\documentclass[letters,useAMS,usenatbib]{mn2e}
\usepackage{graphicx}
\def\gtrsim{\mathrel{\hbox{\rlap{\hbox{\lower4pt\hbox{$\sim$}}}\hbox{$>$}}}}

\begin{document}

\title[Lensing by a Relaxed Galaxy Cluster at z=0.54]{MACSJ1423.8+2404: Gravitational Lensing by a Massive, Relaxed Cluster of Galaxies at z=0.54}
\author[Limousin et al.]{
 \parbox[h]{\textwidth}{
M. Limousin$^{1,2}$
%$\thanks{ased on observations with the NASA/ESA Hubble Space Telescope, obtained
%at the Space Telescope Science Institute, which is operated by the Association of Universities for Research in Astronomy (AURA), Inc. under NASA contract NAS5-26555.}
, H. Ebeling$^{3}$, 
C.-J. Ma$^{3}$, A.~M. Swinbank$^{4}$, G.~P. Smith$^{5,6}$, \\J. Richard$^{4}$,
A.~C. Edge$^{4}$, M. Jauzac$^{1}$, J.-P. Kneib$^{1}$,
P. Marshall$^{7}$ \& T. Schrabback$^{8}$ 
}
\vspace{6pt}\\
$^{1}$ Laboratoire d'Astrophysique de Marseille, Universit\'e de Provence, CNRS, 38 rue Fr\'ed\'eric Joliot-Curie, F-13388 Marseille Cedex
13, France\\
$^{2}$Dark Cosmology Centre, Niels Bohr Institute, University of Copenhagen,
         Juliane Maries Vej 30, 2100 Copenhagen, Denmark\\
$^{3}$Institute for Astronomy, University of Hawaii, 2680 Woodlawn Dr, Honolulu, HI 96822, USA\\
$^{4}$Institute for Computational Cosmology, Department of Physics, Durham
University, South Road, Durham, DH1 3LE, UK\\
$^{5}$School of Physics and Astronomy, University of Birmingham, Edgbaston, Birmingham, B15 2TT, UK\\
$^{6}$California Institute of Technology, Mail Code 105-24, Pasadena, CA, 91125\\
$^{7}$Kavli Institute for Particle Astrophysics and Cosmology, Stanford University, Stanford, CA 94305, USA\\
$^{8}$Leiden Observatory, Leiden University, Niels Bohrweg 2, NL-2333 CA Leiden,
The Netherlands\\
%$^{4}$Jet Propulsion Laboratory, Caltech, MS 169-327, Oak Grove Dr, Pasadena CA 91109, USA\\
}

\date{Accepted. Received; in original form }

\pagerange{\pageref{firstpage}--\pageref{lastpage}} \pubyear{2002}

\maketitle

\label{firstpage}

\begin{abstract}
We present results of a gravitational-lensing and optical study of MACS\,J1423.8+2404 ($z{=}0.545$, MACS\,J1423), the most relaxed cluster in the high-redshift subsample of clusters discovered in the MAssive Cluster Survey (MACS). Our analysis uses high-resolution images taken with the {\it{Hubble Space Telescope}} in the F555W and F814W passbands, ground based imaging in eight optical and near-infrared filters obtained with Subaru and CFHT, as well as extensive spectroscopic data gathered with the Keck telescopes. At optical wavelengths the cluster exhibits no sign of substructure and is dominated by a cD galaxy that is 2.1 magnitudes (K-band) brighter than the second brightest cluster member, suggesting that MACS\,J1423 is close to be fully virialized. Analysis of the redshift distribution of 140 cluster members reveals a Gaussian distribution, mildly disturbed by the presence of a loose galaxy group that may be falling into the cluster along the line of sight. Combining strong-lensing constraints from two spectroscopically confirmed multiple-image systems near the cluster core with a weak-lensing measurement of the gravitational shear on larger scales, we derive a parametric mass model for the mass distribution. All constraints can be satisfied by a uni-modal mass distribution centred on the cD galaxy and exhibiting very little substructure. The derived projected mass of $M(<65\arcsec {\rm [415\,kpc]} )=(4.3\pm0.6)\times 10^{14}$ M$_{\sun}$ is about 30\% higher than the one derived from X-ray analyses assuming spherical symmetry, suggesting a slightly prolate mass distribution consistent with the optical indication of residual line-of-sight structure. The similarity in shape and excellent alignment of the centroids of the total mass, K-band light, and intra-cluster gas distributions add to the picture of a highly evolved system. The existence of a massive cluster like MACS\,J1423, nearly fully virialized only $\sim$ 7 Gyr after the Big Bang, may have important implications for models of structure formation and evolution on cosmological time scales.
\end{abstract}


%________________________________________________________________

\begin{figure*}
\parbox{0.65\textwidth}{
\includegraphics[scale=1.038]{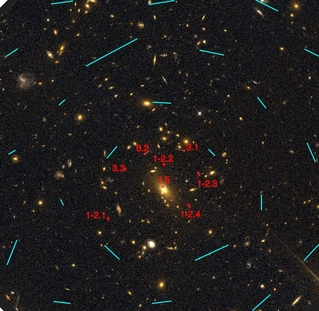}}\hspace*{1.5mm}\mbox{}
\parbox{0.336\textwidth}{
\includegraphics[scale=0.258]{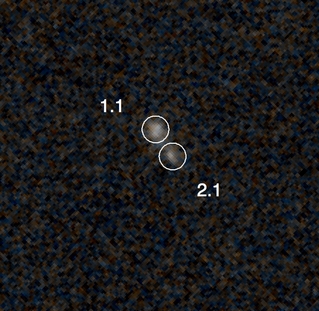}
\includegraphics[scale=0.258]{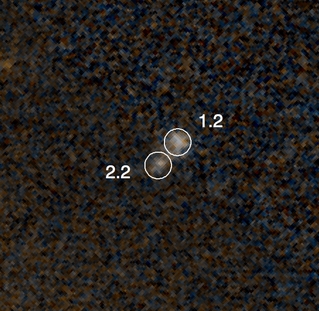}
\includegraphics[scale=0.258]{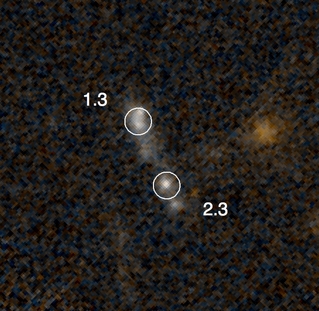}
\includegraphics[scale=0.258]{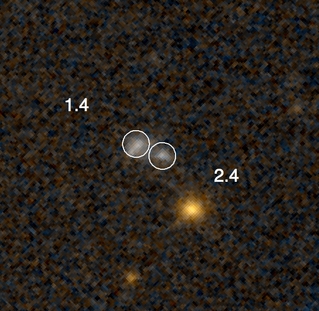}
\includegraphics[scale=0.258]{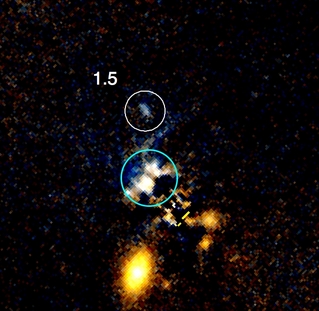}
\includegraphics[scale=0.258]{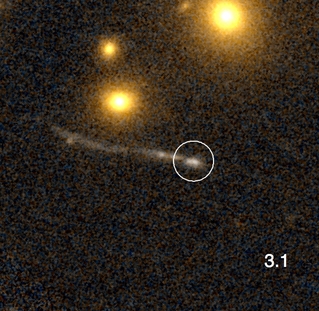}
\includegraphics[scale=0.258]{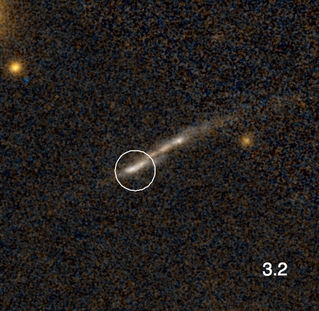}
\includegraphics[scale=0.258]{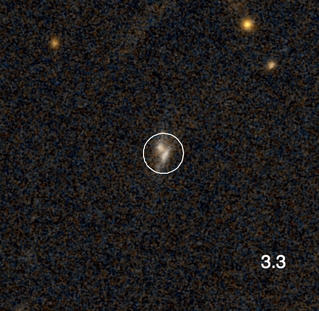}
}
\caption{{\em Left:} MACS\,J1423 as observed with HST/ACS (152$\arcsec\,\times$152$\arcsec$, corresponding to 965$\times$965 kpc$^2$). Multiple-image systems are labelled, the weak-shear field is overlaid in cyan. \emph{Right:} The top four panels (4.7$\arcsec\,\times$4.7$\arcsec$, 30$\times$30 kpc$^2$) show system 1-2, assumed to belong to the same background
source galaxy. System 1 is systematically brighter than system 2. The fifth panel (7.9$\arcsec\,\times$ 7.9$\arcsec$, 50$\times$50 kpc$^2$) shows image 1.5 after the light of the cD galaxy has been subtracted. The cyan circle highlights regions of star formation commonly found in the cD galaxies of cool-core clusters. Indeed, the cD shows strong line emission in IFU observations
(Edge et~al., in prep.), indicative of ongoing cooling. The final three panels show images 3.1, 3.2 and 3.3 respectively (9.4$\arcsec\,\times$9.4$\arcsec$, 60$\times$60 kpc$^2$).
}
\label{fig1}
\end{figure*}

\section{Introduction}
The MAssive Cluster Survey \citep[MACS,][]{macs} has compiled a complete sample of
12 very X-ray luminous galaxy clusters at $z{>}0.5$ \citep{highzmacs}, providing a unique opportunity for comprehensive studies of the densest regions of the cosmic web at intermediate redshifts. As noted by \citet{macslargescale}, MACS\,J1423 stands out among these 12 extreme systems as the most dynamically relaxed; it is, indeed, the most massive cool-core cluster known at these redshifts. The system has been used in several statistical studies \citep{laroque03,1423cosmo,bonamente06,1423MT} where its relaxed dynamical state has been particularly important for cosmological work using the baryon fraction \citep{allen04,schmidtAllen1423,allen08,ettori09}. Since the abundance of massive relaxed clusters at high redshift has important implications for theoretical and numerical models of structure formation and evolution, we here attempt to further test the system's relaxation state using gravitational lensing. All our results use the $\Lambda$CDM concordance cosmology with $\Omega_{\rm{M}} = 0.3, \Omega_\Lambda = 0.7$, and a Hubble constant \textsc{H}$_0 = 70$
km\,s$^{-1}$ Mpc$^{-1}$. Magnitudes are quoted in the AB system.

\section{Observations}

\subsection{Imaging}

Ground based panoramic imaging of MACS\,J1423 was performed in the B, V, R$_c$, I$_c$, and $z'$ bands with the SuprimeCam camera on the Subaru telescope, and in the u$^\ast$ and K bands with MEGACAM and WIRCAM on CFHT. The resulting object catalogues were used to compute photometric redshifts for all galaxies in a $0.5\times 0.5$ deg$^2$ field following the methodology described in \citet{galax0717}. Cluster members were defined to be galaxies with photometric or spectroscopic redshifts (see below) within $\pm$0.05 of the cluster redshift.

MACS\,J1423 was observed on 2004, June 16 with the Advanced Camera for Surveys (ACS)
on-board HST for 4.5 ks and 4.6 ks through the F555W and F814W filters, respectively
(GO: 9722, PI: Ebeling). 
The data were reduced using an automated pipeline developed by the HAGGLeS team (Marshall et al., in prep.), that uses multidrizzle version 2.7.0. Following manual masking of satellite trails, cosmic ray clusters, and scattered light in the flat-fielded frames, all exposures were carefully registered onto a common 0.03 pixel grid \citep{haggles}, generating the shift files needed by multidrizzle. The individual frames in each filter were then drizzled together using a square kernel with pixfrac 0.8, applying updated bad pixel masks and optimal weights \citep{haggles2}.

In spite of the lensing efficiency decreasing with increasing cluster redshift, MACS\,J1423 exhibits clear strong-lensing features. Visual inspection of the ACS frames immediately reveals two multiple-image systems (Fig.~\ref{fig1}). The observed configuration of lensing features for the first of these can be explained by a single background galaxy, resolved into two components, each of which represents a multiple-image system (referred to as 1-2 in the following) in a typical Einstein-cross configuration. The mass model predicts a central fifth image which is detected at the predicted position under the cD galaxy
whose light has been subtracted using the IRAF routine ellipse.
The other multiply imaged system, \#3, constitutes a typical naked-cusp configuration, with three images on the same side of the cluster.

\subsection{Arc spectroscopy}

\begin{figure}
\begin{center}
\includegraphics[scale=0.43,angle=90.0]{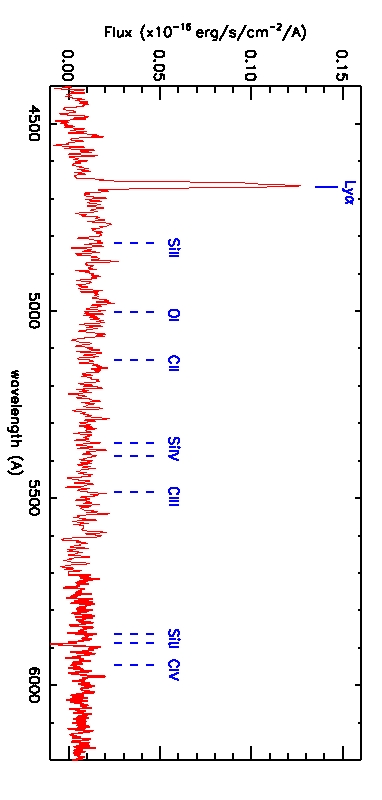}
\includegraphics[scale=0.86,angle=90.0]{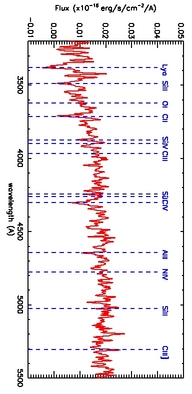}
\caption{Co-added spectra for system 1-2 at $z=2.84$ (\emph{top}) and system 3 at $z=1.779$ (\emph{bottom}).}
\label{spectro}
\end{center}
\end{figure}

Spectroscopic observations of all lensing features were performed on March 6, 2005  using the Low Resolution Imager and Spectrograph \citep[LRIS,][]{lris} of the Keck I telescope in multi-object spectroscopy mode. We used the 400/3400 grism, D560 dichroic, and 400/8500 grating centred at 6800\AA\ to achieve continuous wavelength coverage from 3200 to 8000\AA. The reduction of the data obtained in a total integration time of 10.8\,ks followed the standard steps of bias subtraction, slit identification and extraction, flat fielding, wavelength calibration, and flux calibration, and 
was performed using the \textsc{python} routines described by  \citet{kelson02}.
From the spectra thus obtained we measured redshifts of $z{=}2.84$ for the images labelled 1-2.1, 1-2.2 and 1-2.3; the redshift of features 3.1 and 3.2 was found to be $z=1.779$. 
The spectra for each component show identical features 
(Lyman$\alpha$ in emission for system 1-2, UV absorption lines in system 3),
therefore they have been co-added on Fig.~\ref{spectro}.

\subsection{Galaxy spectroscopy}
\label{velspec} 

\begin{figure}
\includegraphics[scale=0.43]{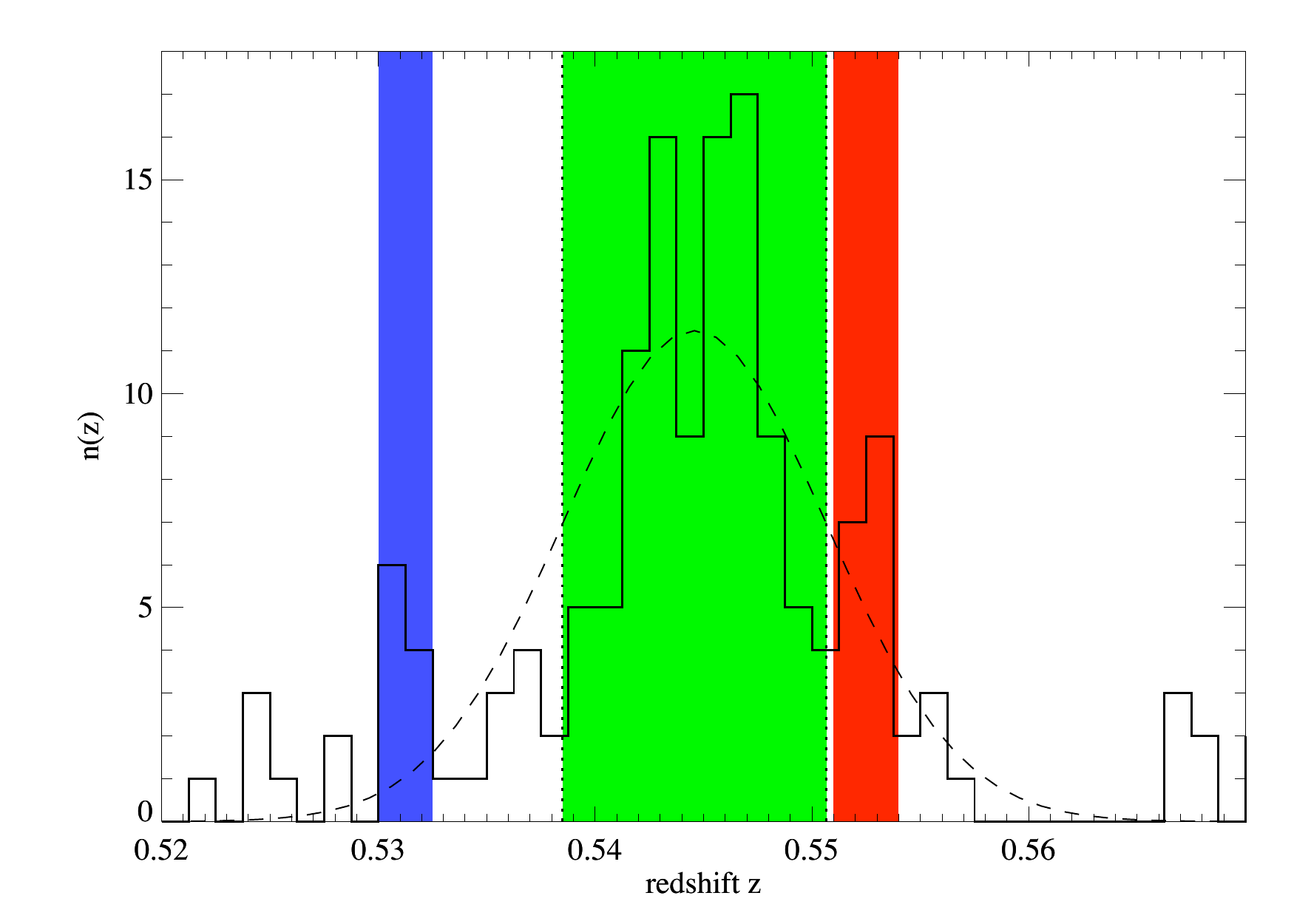}
\includegraphics[scale=0.43]{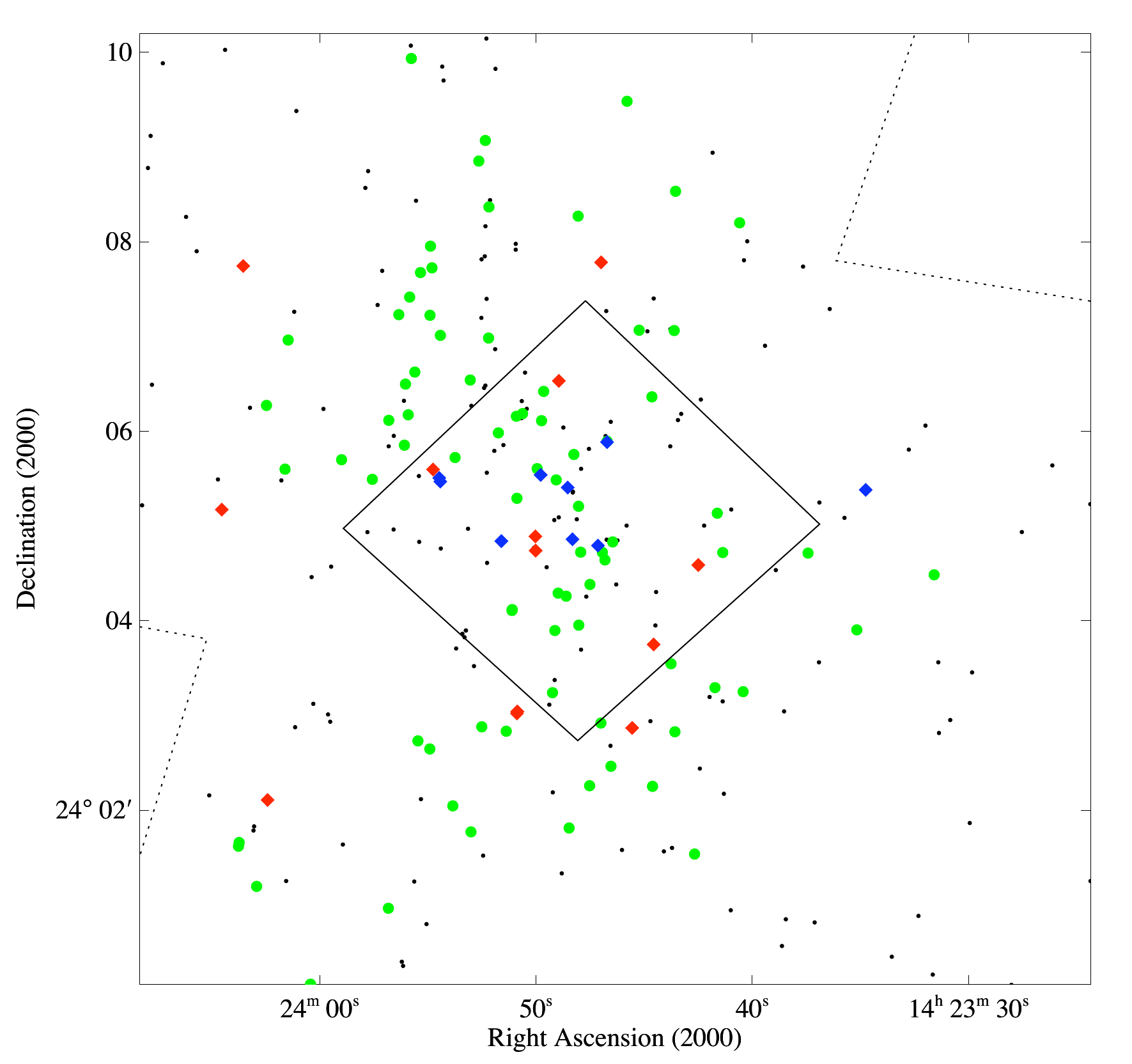}
\caption{
\emph{Top:} Redshift histogram of galaxies in the field of MACS\,J1423. The dashed line shows the best-fitting Gaussian ($z_{\rm cl}{=}0.5450$; $\sigma{=}1320^{+93}_{-91}$ km s$^{-1}$, $n_z{=}140$). The vertical dotted lines mark the $\pm 1\sigma$ range; the coloured bands are designed to separate radial velocities of potential fore- or background structure.
\emph{Bottom:} The distribution of selected cluster members on the sky (same colour scheme as in top figure: the main cluster galaxies are shown as green
circles, and the populations of the two peaks are shown as filled diamonds). 
Galaxies with $z\sim 0.531$ (blue) are tightly concentrated and thus likely to constitute a bound galaxy group, falling into the cluster core along our line of sight. The black square outlines the area covered by our HST/ACS observation.
The dashed lines delineate
small areas that are not covered by our spectroscopic survey. The small black
dots represent other galaxies with spectroscopic redshifts but outside the
cluster.
}
\label{veldist}
\end{figure}

Using the LRIS and DEIMOS spectrographs on the twin Keck-10m telescopes on Mauna Kea we conducted an extensive spectroscopic survey of galaxies in the field of MACS\,J1423. Redshifts were measured for 396 objects, 60 of them within the solid angle covered by our HST/ACS observation. All redshifts were used to calibrate the photometric redshifts mentioned above. Using iterative $3\sigma$ clipping we find $z_{\rm cl}{=}0.5450$ and $\sigma{=}1320^{+93}_{-91}$ km s$^{-1}$ for the cluster redshift and velocity dispersion, respectively ($n_{\rm z}{=}140$, Fig.~\ref{veldist}). When limited to the ACS field of view, the same analysis yields $z_{\rm cl}{=}0.5424$ and $\sigma{=}1470^{+170}_{-120}$ km s$^{-1}$ ($n_{\rm z}{=}40$). Even for a system as 
X-ray luminous as MACS\,J1423 \citep[$L_{\rm X, bol}=3.7\times 10^{45}$ erg s$^{-1}$,][]{highzmacs}, this velocity dispersion falls high on the $L_{\rm X}$-$\sigma$ relation \citep[e.g.][]{mahdavi01}, without, however, constituting a significant outlier. 

Inspection of the galaxy velocity histogram (Fig.~\ref{veldist}, top) shows the distribution to be mildly 
skewed toward lower redshifts; a one-sided Kolmogorov-Smirnov test finds it to be inconsistent with a Gaussian at the 2$\sigma$ confidence level. We investigate whether the two peaks at $z{\sim}0.531$ and $z{\sim}0.553$ represent subclusters along the line of sight by selecting three galaxy subsamples as indicated by the coloured bands. The distribution of the respective galaxies on the sky is shown in Fig.~\ref{veldist} (bottom). Unlike the galaxies with $z{\sim}0.553$, we find the galaxies in the peak around $z{\sim}0.531$ to be significantly (3.3$\sigma$) more concentrated in projection on the sky than the overall cluster galaxy population, suggesting a gravitationally bound substructure viewed almost exactly along the line of sight. We thus conclude that the velocity dispersion of MACS\,J1423 is mildly inflated by a possible infall of a loose group at $z=0.531$, viewed in projection and superimposed directly onto the cluster core. 

\section{Distribution of Light \& Mass}

\subsection{Light Distribution}
We use the K-band magnitudes of (photometric) cluster members to create a luminosity-density map of the cluster light. By fitting a  2D Sersic profile to the cD
galaxy which dominates the K-band cluster light (we use the \textsc{galfit} software of \citet{galfit}), we measure an integrated magnitude of K{=}16.27 for the cD, compared to  K{=}18.33 for the second brightest cluster member. The magnitude difference is
K$_{\rm gap}{=}2.06$.

\subsection{Lensing Analysis}

In addition to the two spectroscopically confirmed strong-lensing features discussed above, we also take into account the weak-lensing signal as reflected in the sheared F814W images of the 
background galaxy population. Details of the weak-lensing analysis will be presented 
in a forthcoming publication (Jauzac et al., in prep.).
In short, we follow the weak lensing pipeline developped by
\citet{alexie1,alexie2}, but adapted the code to work on cluster field,
in order to efficiently remove the galaxy cluster members and the foreground
galaxies that would otherwise reduce the strength of the shear signal.
In doing so, we consider only background sources located outside the Einstein radius 
of, for MACS\,J1423, $R_{\rm E}=21\arcsec$ (here we refer to the effective Einstein 
for a source at $z=2$, defined as the angular radius from the centre of the cluster
at which the average convergence equals 1), thereby minimising any systematic bias from the strong-shear 
regime \citep{masseySL}. The 192 galaxies thus selected within the ACS 
field (corresponding to a number density of 16 galaxies per
square arcminute) represent 192 constraints in the weak-lensing analysis. Adding the 18 constraints from the two strongly lensed multiple-image systems yields a total of 210 constraints, all of which are used in order to build a parametric mass distribution within the framework of the publicly available \textsc{lenstool} software 
\citep[][http://www.oamp.fr/cosmology/lenstool/]{jullo07}
in which we have implemented the inclusion of the weak lensing constraints. The optimisation is performed in the image plane.
 
To characterise the mass distribution in MACS\,J1423 we adopt a dual Pseudo Isothermal Elliptical
Mass Distribution \citep[dPIE,][]{mypaperI,ardis2218}, parametrised by a fiducial velocity dispersion $\sigma$, a core radius $r_{\rm core}$ and a scale radius $r_s$ (set to 1\,000$\arcsec$). In addition to this large-scale mass component,  which models both the intracluster  gas and the far larger dark matter component \citep{bradac08a,jesper}, we also take into account the perturbations associated with individual cluster galaxies. Specifically, we include in the modelling all  cluster members brighter than 22.5 (F814) and within a projected cluster-centric distance of 35$\arcsec$ from the cD galaxy. Empirical scaling relations (without any scatter) are used to relate their dynamical dPIE parameters (central velocity dispersion and scale radius) to their luminosity (the core radius being set to 0), whereas all geometrical parameters (centre, ellipticity, position angle) are set to the values measured from the light distribution
\citep[see, \emph{e.g.}][]{mypaperIII,a370}.
In total our mass model is thus described by eight free parameters.
The range of allowed parameter values characterising the dPIE mass component is limited by priors as follows. Position:  $\pm\, 25\arcsec$ along the X and Y directions; ellipticity: ${<}0.6$ (expressed in units $(a^2-b^2)/(a^2+b^2)$, where $a$ and $b$ refers 
to the semi-major and semi-minor axes of the ellipse describing the mass distribution); position angle: 0--180 degrees; velocity dispersion: 500--1500 km$\,s^{-1}$; core radius: 1--20$\arcsec$. For the galaxy-scale component, we allow the velocity dispersion to vary between 100 and 250 km\,s$^{-1}$, whereas the scale radius was forced to be less than 70 kpc in order to account for tidal stripping of their dark matter haloes
\citep[see, \emph{e.g.}][and references therein.]{mypaperII,mypaperIV,priya4,wetzel} 

\subsection{A Uni-modal Mass Distribution}

\begin{table*}
\begin{center}
\begin{tabular}{ccccccc}
\hline
Component & $\delta x (\arcsec)$ & $\delta y (\arcsec)$ & $e$ & $\theta$ & $r$ (\footnotesize{kpc}) & $\sigma$ (\footnotesize{km\,s$^{-1}$})\\
\hline
\smallskip
\smallskip
Cluster Halo &  -1.0$\pm$1.2 & 0.0$\pm$1.0 & 0.4$\pm$0.1 & 116.0$\pm$1.7 & 78.7$\pm$25.2 & 1081$\pm$110 \\
\smallskip
Perturbers & - & - & - & - & 32$^{+66}_{-30}$ & 141$^{+54}_{-21}$ 
\smallskip
\end{tabular}
\end{center}
\caption{Best-fit values of all mass-model parameters.
Coordinates are defined with respect to the cD galaxy.
Error bars correspond
to $3\sigma$ confidence level as inferred from the \textsc{mcmc} optimisation. 
%When the posterior probability distribution is not Gaussian, we report the mode and asymmetric error bars.
For the dPIE mass component, $r$ corresponds to the core radius whereas for the galaxy-scale perturbations it corresponds to the scale radius.
\label{restable}}
\end{table*}

The shear field constructed from our analysis of the images of the weakly lensed background galaxy population (Fig.~\ref{fig1}) already suggests a uni-modal mass distribution centred on the
cD galaxy. Combining strong- and weak-lensing constraints, we confirm that a 
single cluster-scale mass component is able to reproduce the full set of observational
constraints, with a reduced $\chi^2$ value of 175/202.
The corresponding best-fit values of all model parameters are summarised in Table~1. We find the centre of the elliptical dPIE mass distribution to coincide with that of the cD galaxy; its position angle of 116$\pm$2 degrees matches the orientation of the cD galaxy (126$\pm$10 degrees). No additional large-scale mass component is required by the data.

\begin{figure}
\begin{center}
\includegraphics[scale=0.43]{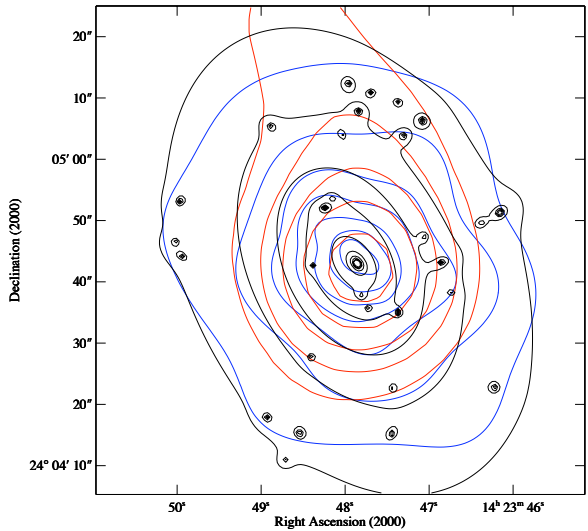}
\caption{The distribution of mass, gas, and galaxies in the core of MACS\,J1423, as reflected
by the adaptively smoothed K-band luminosity of cluster galaxies (red), the total
mass derived from the gravitational-lens model (black) and the adaptively smoothed X-ray
surface brightness from \emph{Chandra} observations (blue). 
}
\label{fig2}
\end{center}
\end{figure}

Fig.~\ref{fig2} compares the 2D mass map corresponding to the best-fit lens model to the X-ray surface brightness as observed with \emph{Chandra} 
(from \citet{highzmacs})
and to the K-band light map, both adaptively smoothed to 3$\sigma$ significance using \textsc{asmooth} \citep{asmooth}. The agreement between these independent observational tracers of the three cluster constituents (dark matter, gas, and galaxies) is remarkable. 

%The mean projected mass enclosed within a given aperture, as estimated from realisations using the MCMC sampler, is plotted in Fig.~\ref{mass}.

%\begin{figure}
%\begin{center}
%\includegraphics[scale=0.5]{f4.jpg}
%\caption{Total mass as a function of projected radius. The solid lines delineate the 3$\sigma$ confidence limits to the lensing mass as derived from the analysis presented here. For comparison we show independent mass estimates at R$= 65\arcsec$ from \citet{schmidtAllen1423} (solid) and \citet{laroque03} (dashed). The squares show the radial location of the strong-lensing constraints. 
%}
%\label{mass}
%\end{center}
%\end{figure}

\section{Discussion \& Conclusions}

Using optical and lensing observations of MACS\,J1423 we find overwhelming evidence for a uni-modal mass distribution whose centre, ellipticity, and orientation are in excellent
agreement with the corresponding parameters of the cluster light and intra-cluster gas distributions (Fig.~\ref{fig2}). We quantify the dominance of a single, central mass component by computing the substructure fraction, $f_{\rm sub}$, defined as the mass within a given cluster-centric radius that can be attributed to substructure, divided by the total mass within the same radius \citep{smithtaylor}. In the case of MACS\,J1423, only cluster galaxies (the cD galaxy is by definition excluded) contribute to the substructure fraction.  
We find $f_{\rm sub}(<250\,{\rm kpc}){=}0.048 \pm$0.025, comparable to values measured 
for the most relaxed clusters at $z\sim0.2$ \citep{locuss}. Within the high-redshift MACS subsample, the substructure fraction of $f_{\rm sub}(<500\,{\rm kpc}){=}0.045\pm$0.025 of MACS\,J1423 is in stark contrast to the value of  0.25$\pm$0.12 measured for MACS\,J1149.5+2223, the most complex cluster lens studied to date \citep{1149}.

The radial mass profile of MACS\,J1423 as derived from our lensing analysis is well constrained across the full ACS field of view, i.e., out to a radius of 100\arcsec (640\,kpc). Previous results for the mass within 65\arcsec (415\,kpc) of 5.0$^{+3.1}_{-0.9}\times 10^{14}$ M$_{\sun}$, \citep{laroque03}\footnote{\citet{laroque03} quote the mass within a three-dimensional radius, which we convert to a projected mass assumed a $\beta$-model with $b=2/3$.} using observations of the Sun\-yaev-Zel`dovich effect, and of 3.1$^{+0.9}_{-0.7} \times 10^{14}$ M$_{\sun}$ \citep[3$\sigma$ confidence level,][]{schmidtAllen1423} from an analysis of the {\em Chandra} X-ray data, agree within the errors with our measurement of $4.35\pm0.6 \times 10^{14}$ M$_{\sun}$ (3$\sigma$). We note, however, that the existing X-ray analyses of MACS\,J1423 assume spherical symmetry which will bias the resulting mass estimates low for a prolate matter distribution. The difference of $\sim$ 30\% between the lensing and X-ray mass may thus be due to an elongation 
of the system along the line of sight, consistent with our finding of residual accretion onto the cluster core (Section~\ref{velspec}).

\citet{locuss} investigated the relationship between the effective Einstein
radius and the strong lensing mass (measured within 250 kpc) for a sample
of 20 strong lensing clusters at $z\sim0.2$ drawn from the Local Cluster
Substructure Survey (LoCuSS). According to their statistical analysis, one would expect
a mass of 2.55$^{+0.90}_{-0.65}$ 10$^{14}$ M$_{\sun}$ for an \emph{undisturbed}
cluster with an Einstein radius of 21$\arcsec$. 
The excellent agreement with our measurement of 2.46$\pm$0.2 10$^{14}$ M$_{\sun}$
underlines again the relaxed dynamical state of MACS\,J1423.

Although our analysis of the radial velocity distribution of cluster members suggests that MACS\,J1423 evolved through mergers along a line-of-sight filament (Section~\ref{velspec}), the rate of current or recent infall is low and does not result in a noticeable perturbation of the cluster core. In fact, the large luminosity gap of K$_{\rm gap}{=}2.06$ mag between the cD and the second brightest cluster galaxy, as well as the excellent match of the dark-matter, intra-cluster gas, and cluster light distributions, represent strong evidence for a dynamically highly evolved cluster core \citep{jones03,Lgapcluster}, undisturbed by major mergers for at least 1 Gyr, the characteristic time scale for relaxation via dynamical friction. The existence of  MACS\,J1423 only 7 Gyr after the Big Bang thus adds an important observational constraint to N-body simulations of structure formation and evolution in the current cosmological model.

\section*{Acknowledgements}
ML thanks the Centre National d'Etudes Spatiales (CNES) and CNRS for their support.
ML est b\'en\'eficiaire d'une bourse d'acceuil de la Ville de Marseille.
The Dark Cosmology Centre is funded by the Danish National National
Research Foundation.
HE gratefully acknowledges financial support from STScI grant GO-09722 and SAO grant GO3-4164X.
AMS gratefully acknowledges a Royal Astronomical Society Sir Norman Lockyer Fellowship.
GPS acknowledges support from the Royal Society and STFC.
ACE and IRS acknowledge support from STFC.
JR acknowledges support from an EU Marie-Curie fellowship.
JPK acknowledges support from CNRS.
TS acknowledges support from the Netherlands Organization for Scientific Research (NWO).

%\bibliography{references}
\bibliography{draft}

\begin{thebibliography}{38}
\expandafter\ifx\csname natexlab\endcsname\relax\def\natexlab#1{#1}\fi

\bibitem[{{Allen} {et~al.}(2008){Allen}, {Rapetti}, {Schmidt}, {Ebeling},
  {Morris}, \& {Fabian}}]{allen08}
{Allen} S.~W., {Rapetti} D.~A., {Schmidt} R.~W., {Ebeling} H., {Morris} R.~G.,
  {Fabian} A.~C., 2008, \mnras, 383, 879

\bibitem[{{Allen} {et~al.}(2004){Allen}, {Schmidt}, {Ebeling}, {Fabian}, \&
  {van Speybroeck}}]{allen04}
{Allen} S.~W., {Schmidt} R.~W., {Ebeling} H., {Fabian} A.~C., {van Speybroeck}
  L., 2004, \mnras, 353, 457

\bibitem[{{Bonamente} {et~al.}(2006){Bonamente}, {Joy}, {LaRoque}, {Carlstrom},
  {Reese}, \& {Dawson}}]{bonamente06}
{Bonamente} M., {Joy} M.~K., {LaRoque} S.~J., {Carlstrom} J.~E., {Reese} E.~D.,
  {Dawson} K.~S., 2006, \apj, 647, 25

\bibitem[{{Brada{\v c}} {et~al.}(2008){Brada{\v c}}, {Schrabback}, {Erben},
  {McCourt}, {Million}, {Mantz}, {Allen}, {Blandford}, {Halkola},
  {Hildebrandt}, {Lombardi}, {Marshall}, {Schneider}, {Treu}, \&
  {Kneib}}]{bradac08a}
{Brada{\v c}} M., {Schrabback} T., {Erben} T., {McCourt} M., {Million} E.,
  {Mantz} A., {Allen} S., {Blandford} R., {Halkola} A., {Hildebrandt} H.,
  {Lombardi} M., {Marshall} P., {Schneider} P., {Treu} T., {Kneib} J.-P., 2008,
  \apj, 681, 187

\bibitem[{{Ebeling} {et~al.}(2007){Ebeling}, {Barrett}, {Donovan}, {Ma},
  {Edge}, \& {van Speybroeck}}]{highzmacs}
{Ebeling} H., {Barrett} E., {Donovan} D., {Ma} C.-J., {Edge} A.~C., {van
  Speybroeck} L., 2007, \apjl, 661, L33

\bibitem[{{Ebeling} {et~al.}(2001){Ebeling}, {Edge}, \& {Henry}}]{macs}
{Ebeling} H., {Edge} A.~C., {Henry} J.~P., 2001, \apj, 553, 668

\bibitem[{{Ebeling} {et~al.}(2006){Ebeling}, {White}, \&
  {Rangarajan}}]{asmooth}
{Ebeling} H., {White} D.~A., {Rangarajan} F.~V.~N., 2006, \mnras, 368, 65

\bibitem[{{El{\'{\i}}asd{\'o}ttir} {et~al.}(2007){El{\'{\i}}asd{\'o}ttir},
  {Limousin}, {Richard}, {Hjorth}, {Kneib}, {Natarajan}, {Pedersen}, {Jullo},
  \& {Paraficz}}]{ardis2218}
{El{\'{\i}}asd{\'o}ttir} {\'A}., {Limousin} M., {Richard} J., {Hjorth} J.,
  {Kneib} J.-P., {Natarajan} P., {Pedersen} K., {Jullo} E., {Paraficz} D.,
  2007, ArXiv e-prints, 710

\bibitem[{{Ettori} {et~al.}(2009){Ettori}, {Morandi}, {Tozzi}, {Balestra},
  {Borgani}, {Rosati}, {Lovisari}, \& {Terenziani}}]{ettori09}
{Ettori} S., {Morandi} A., {Tozzi} P., {Balestra} I., {Borgani} S., {Rosati}
  P., {Lovisari} L., {Terenziani} F., 2009, ArXiv e-prints

\bibitem[{{Jones} {et~al.}(2003){Jones}, {Ponman}, {Horton}, {Babul},
  {Ebeling}, \& {Burke}}]{jones03}
{Jones} L.~R., {Ponman} T.~J., {Horton} A., {Babul} A., {Ebeling} H., {Burke}
  D.~J., 2003, \mnras, 343, 627

\bibitem[{{Jullo} {et~al.}(2007){Jullo}, {Kneib}, {Limousin},
  {El{\'{\i}}asd{\'o}ttir}, {Marshall}, \& {Verdugo}}]{jullo07}
{Jullo} E., {Kneib} J.-P., {Limousin} M., {El{\'{\i}}asd{\'o}ttir} {\'A}.,
  {Marshall} P.~J., {Verdugo} T., 2007, New Journal of Physics, 9, 447

\bibitem[{{Kartaltepe} {et~al.}(2008){Kartaltepe}, {Ebeling}, {Ma}, \&
  {Donovan}}]{macslargescale}
{Kartaltepe} J.~S., {Ebeling} H., {Ma} C.~J., {Donovan} D., 2008, \mnras, 389,
  1240

\bibitem[{{Kelson} {et~al.}(2002){Kelson}, {Zabludoff}, {Williams}, {Trager},
  {Mulchaey}, \& {Bolte}}]{kelson02}
{Kelson} D.~D., {Zabludoff} A.~I., {Williams} K.~A., {Trager} S.~C., {Mulchaey}
  J.~S., {Bolte} M., 2002, \apj, 576, 720

\bibitem[{{Kotov} \& {Vikhlinin}(2006)}]{1423MT}
{Kotov} O., {Vikhlinin} A., 2006, \apj, 641, 752

\bibitem[{{LaRoque} {et~al.}(2003){LaRoque}, {Joy}, {Carlstrom}, {Ebeling},
  {Bonamente}, {Dawson}, {Edge}, {Holzapfel}, {Miller}, {Nagai}, {Patel}, \&
  {Reese}}]{laroque03}
{LaRoque} S.~J., {Joy} M., {Carlstrom} J.~E., {Ebeling} H., {Bonamente} M.,
  {Dawson} K.~S., {Edge} A., {Holzapfel} W.~L., {Miller} A.~D., {Nagai} D.,
  {Patel} S.~K., {Reese} E.~D., 2003, \apj, 583, 559

\bibitem[{{Leauthaud} {et~al.}(2010){Leauthaud}, {Finoguenov}, {Kneib},
  {Taylor}, {Massey}, {Rhodes}, {Ilbert}, {Bundy}, {Tinker}, {George}, {Capak},
  {Koekemoer}, {Johnston}, {Zhang}, {Cappelluti}, {Ellis}, {Elvis}, {Giodini},
  {Heymans}, {Le F{\`e}vre}, {Lilly}, {McCracken}, {Mellier},
  {R{\'e}fr{\'e}gier}, {Salvato}, {Scoville}, {Smoot}, {Tanaka}, {Van
  Waerbeke}, \& {Wolk}}]{alexie2}
{Leauthaud} A., {Finoguenov} A., {Kneib} J., {Taylor} J.~E., {Massey} R.,
  {Rhodes} J., {Ilbert} O., {Bundy} K., {Tinker} J., {George} M.~R., {Capak}
  P., {Koekemoer} A.~M., {Johnston} D.~E., {Zhang} Y., {Cappelluti} N., {Ellis}
  R.~S., {Elvis} M., {Giodini} S., {Heymans} C., {Le F{\`e}vre} O., {Lilly} S.,
  {McCracken} H.~J., {Mellier} Y., {R{\'e}fr{\'e}gier} A., {Salvato} M.,
  {Scoville} N., {Smoot} G., {Tanaka} M., {Van Waerbeke} L., {Wolk} M., 2010,
  \apj, 709, 97

\bibitem[{{Leauthaud} {et~al.}(2007){Leauthaud}, {Massey}, {Kneib}, {Rhodes},
  {Johnston}, {Capak}, {Heymans}, {Ellis}, {Koekemoer}, {Le F{\`e}vre},
  {Mellier}, {R{\'e}fr{\'e}gier}, {Robin}, {Scoville}, {Tasca}, {Taylor}, \&
  {Van Waerbeke}}]{alexie1}
{Leauthaud} A., {Massey} R., {Kneib} J., {Rhodes} J., {Johnston} D.~E., {Capak}
  P., {Heymans} C., {Ellis} R.~S., {Koekemoer} A.~M., {Le F{\`e}vre} O.,
  {Mellier} Y., {R{\'e}fr{\'e}gier} A., {Robin} A.~C., {Scoville} N., {Tasca}
  L., {Taylor} J.~E., {Van Waerbeke} L., 2007, \apjs, 172, 219

\bibitem[{{Limousin} {et~al.}(2007{\natexlab{a}}){Limousin}, {Kneib},
  {Bardeau}, {Natarajan}, {Czoske}, {Smail}, {Ebeling}, \& {Smith}}]{mypaperII}
{Limousin} M., {Kneib} J.~P., {Bardeau} S., {Natarajan} P., {Czoske} O.,
  {Smail} I., {Ebeling} H., {Smith} G.~P., 2007{\natexlab{a}}, \aap, 461, 881

\bibitem[{{Limousin} {et~al.}(2005){Limousin}, {Kneib}, \&
  {Natarajan}}]{mypaperI}
{Limousin} M., {Kneib} J.-P., {Natarajan} P., 2005, \mnras, 356, 309

\bibitem[{{Limousin} {et~al.}(2007{\natexlab{b}}){Limousin}, {Richard},
  {Jullo}, {Kneib}, {Fort}, {Soucail}, {El{\'{\i}}asd{\'o}ttir}, {Natarajan},
  {Ellis}, {Smail}, {Czoske}, {Smith}, {Hudelot}, {Bardeau}, {Ebeling},
  {Egami}, \& {Knudsen}}]{mypaperIII}
{Limousin} M., {Richard} J., {Jullo} E., {Kneib} J.~P., {Fort} B., {Soucail}
  G., {El{\'{\i}}asd{\'o}ttir} A., {Natarajan} P., {Ellis} R.~S., {Smail} I.,
  {Czoske} O., {Smith} G.~P., {Hudelot} P., {Bardeau} S., {Ebeling} H., {Egami}
  E., {Knudsen} K.~K., 2007{\natexlab{b}}, \apj, 668, 643

\bibitem[{{Limousin} {et~al.}(2009){Limousin}, {Sommer-Larsen}, {Natarajan}, \&
  {Milvang-Jensen}}]{mypaperIV}
{Limousin} M., {Sommer-Larsen} J., {Natarajan} P., {Milvang-Jensen} B., 2009,
  \apj, 696, 1771

\bibitem[{{Ma} {et~al.}(2008){Ma}, {Ebeling}, {Donovan}, \&
  {Barrett}}]{galax0717}
{Ma} C.-J., {Ebeling} H., {Donovan} D., {Barrett} E., 2008, \apj, 684, 160

\bibitem[{{Mahdavi} \& {Geller}(2001)}]{mahdavi01}
{Mahdavi} A., {Geller} M.~J., 2001, \apjl, 554, L129

\bibitem[{{Massey} \& {Goldberg}(2008)}]{masseySL}
{Massey} R., {Goldberg} D.~M., 2008, \apjl, 673, L111

\bibitem[{{Milosavljevi{\'c}} {et~al.}(2006){Milosavljevi{\'c}}, {Miller},
  {Furlanetto}, \& {Cooray}}]{Lgapcluster}
{Milosavljevi{\'c}} M., {Miller} C.~J., {Furlanetto} S.~R., {Cooray} A., 2006,
  \apjl, 637, L9

\bibitem[{{Natarajan} {et~al.}(2009){Natarajan}, {Kneib}, {Smail}, {Treu},
  {Ellis}, {Moran}, {Limousin}, \& {Czoske}}]{priya4}
{Natarajan} P., {Kneib} J.-P., {Smail} I., {Treu} T., {Ellis} R., {Moran} S.,
  {Limousin} M., {Czoske} O., 2009, \apj, 693, 970

\bibitem[{{Oke} {et~al.}(1995){Oke}, {Cohen}, {Carr}, {Cromer}, {Dingizian},
  {Harris}, {Labrecque}, {Lucinio}, {Schaal}, {Epps}, \& {Miller}}]{lris}
{Oke} J.~B., {Cohen} J.~G., {Carr} M., {Cromer} J., {Dingizian} A., {Harris}
  F.~H., {Labrecque} S., {Lucinio} R., {Schaal} W., {Epps} H., {Miller} J.,
  1995, \pasp, 107, 375

\bibitem[{{Peng} {et~al.}(2002){Peng}, {Ho}, {Impey}, \& {Rix}}]{galfit}
{Peng} C.~Y., {Ho} L.~C., {Impey} C.~D., {Rix} H.-W., 2002, \aj, 124, 266

\bibitem[{{Puetzfeld} {et~al.}(2005){Puetzfeld}, {Pohl}, \& {Zhu}}]{1423cosmo}
{Puetzfeld} D., {Pohl} M., {Zhu} Z.-H., 2005, \apj, 619, 657

\bibitem[{{Richard} {et~al.}(2009{\natexlab{a}}){Richard}, {Kneib}, {Limousin},
  {Edge}, \& {Jullo}}]{a370}
{Richard} J., {Kneib} J., {Limousin} M., {Edge} A., {Jullo} E.,
  2009{\natexlab{a}}, ArXiv e-prints

\bibitem[{{Richard} {et~al.}(2009{\natexlab{b}}){Richard}, {Smith}, {Kneib},
  {Ellis}, {Sanderson}, {Pei}, {Targett}, {Sand}, {Swinbank}, {Dannerbauer},
  {Mazzotta}, {Limousin}, {Egami}, {Jullo}, {Hamilton-Morris}, \&
  {Moran}}]{locuss}
{Richard} J., {Smith} G., {Kneib} J., {Ellis} R., {Sanderson} A., {Pei} L.,
  {Targett} T., {Sand} D., {Swinbank} M., {Dannerbauer} H., {Mazzotta} P.,
  {Limousin} M., {Egami} E., {Jullo} E., {Hamilton-Morris} V., {Moran} S.,
  2009{\natexlab{b}}, ArXiv e-prints

\bibitem[{{Schmidt} \& {Allen}(2007)}]{schmidtAllen1423}
{Schmidt} R.~W., {Allen} S.~W., 2007, \mnras, 379, 209

\bibitem[{{Schrabback} {et~al.}(2007){Schrabback}, {Erben}, {Simon},
  {Miralles}, {Schneider}, {Heymans}, {Eifler}, {Fosbury}, {Freudling},
  {Hetterscheidt}, {Hildebrandt}, \& {Pirzkal}}]{haggles}
{Schrabback} T., {Erben} T., {Simon} P., {Miralles} J., {Schneider} P.,
  {Heymans} C., {Eifler} T., {Fosbury} R.~A.~E., {Freudling} W.,
  {Hetterscheidt} M., {Hildebrandt} H., {Pirzkal} N., 2007, \aap, 468, 823

\bibitem[{{Schrabback} {et~al.}(2009){Schrabback}, {Hartlap}, {Joachimi},
  {Kilbinger}, {Simon}, {Benabed}, {Brada{\v c}}, {Eifler}, {Erben},
  {Fassnacht}, {High}, {Hilbert}, {Hildebrandt}, {Hoekstra}, {Kuijken},
  {Marshall}, {Mellier}, {Morganson}, {Schneider}, {Semboloni}, {Van Waerbeke},
  \& {Velander}}]{haggles2}
{Schrabback} T., {Hartlap} J., {Joachimi} B., {Kilbinger} M., {Simon} P.,
  {Benabed} K., {Brada{\v c}} M., {Eifler} T., {Erben} T., {Fassnacht} C.~D.,
  {High} F.~W., {Hilbert} S., {Hildebrandt} H., {Hoekstra} H., {Kuijken} K.,
  {Marshall} P., {Mellier} Y., {Morganson} E., {Schneider} P., {Semboloni} E.,
  {Van Waerbeke} L., {Velander} M., 2009, ArXiv e-prints

\bibitem[{{Smith} {et~al.}(2009){Smith}, {Ebeling}, {Limousin}, {Kneib},
  {Swinbank}, {Ma}, {Jauzac}, {Richard}, {Jullo}, {Sand}, {Edge}, \&
  {Smail}}]{1149}
{Smith} G.~P., {Ebeling} H., {Limousin} M., {Kneib} J., {Swinbank} A.~M., {Ma}
  C., {Jauzac} M., {Richard} J., {Jullo} E., {Sand} D.~J., {Edge} A.~C.,
  {Smail} I., 2009, ArXiv e-prints

\bibitem[{{Smith} \& {Taylor}(2008)}]{smithtaylor}
{Smith} G.~P., {Taylor} J.~E., 2008, \apjl, 682, L73

\bibitem[{{Sommer-Larsen} \& {Limousin}(2009)}]{jesper}
{Sommer-Larsen} J., {Limousin} M., 2009, ArXiv e-prints

\bibitem[{{Wetzel} \& {White}(2009)}]{wetzel}
{Wetzel} A.~R., {White} M., 2009, ArXiv e-prints

\end{thebibliography}

\label{lastpage}

\end{document}